\documentclass[twocolumn,superscriptaddress,altaffilletter,lengthcheck,showpacs, tightenlines,showkeys, prd]{revtex4-1}
\usepackage{amssymb}
\usepackage{amsfonts}
\usepackage{subcaption}
\usepackage{graphicx}
\usepackage{caption}
\usepackage{xcolor}
\usepackage{amsmath}
\usepackage{amstext}
 \usepackage[dvipdf]{epsfig}
  \usepackage{color}
\def\be{\begin{equation}}
\def\ee#1{\label{#1}\end{equation}}
\newcommand{\ben}{\begin{eqnarray}}
\newcommand{\een}{\end{eqnarray}}

\begin{document}
\title{Quasinormal modes and self-adjoint extensions of the Schroedinger operator}
\author{J\'ulio C. Fabris}\email{julio.fabris@cosmo-ufes.org}
\affiliation{PPGCosmo, CCE - Universidade Federal do Esp\'irito Santo,  29075-910 Vit\'oria, ES, Brazil}
\affiliation{N\'ucleo Cosmo-ufes\&Departamento de F\'isica - Universidade Federal do Esp\'irito Santo,   29075-910  Vit\'oria, ES, Brazil}
\affiliation{National Research Nuclear University MEPhI, Kashirskoe sh. 31, Moscow 115409, Russia}
\author{Mart\'{\i}n G. Richarte}\email{martin@df.uba.ar  }
\affiliation{PPGCosmo, CCE - Universidade Federal do Esp\'irito Santo,  29075-910 Vit\'oria, ES, Brazil}
\affiliation{Departamento de F\'isica, Facultad de Ciencias Exactas y Naturales,
Universidad de Buenos Aires, Ciudad Universitaria 1428, Pabell\'on I,  Buenos Aires, Argentina}
\author{Alberto Saa}\email{asaa@ime.unicamp.br}
\affiliation{Departamento de Matem\'atica Aplicada, Universidade Estadual de Campinas, 13083-859 Campinas, SP, Brazil}
\bibliographystyle{plain}

\begin{abstract}
We revisit here the analytical continuation approach usually employed to compute quasinormal modes (QNM) and frequencies of a
given potential barrier $V$ starting from the  bounded states 
and   respective eigenvalues 
of the Schroedinger operator 
associated with 
the 
potential well corresponding to the 
inverted potential   $-V$.  We consider an exactly soluble problem corresponding to a potential barrier of the Poschl-Teller type
with a well defined and behaved QNM spectrum, but for which the associated  Schroedinger operator  $\cal H$ 
obtained by   analytical continuation fails to be self-adjoint. Although  $\cal H$ 
admits self-adjoint extensions, we  show that 
the  eigenstates   corresponding to the analytically continued   QNM  
do not belong to any self-adjoint extension domain  and, consequently, they cannot be interpreted as
authentic  quantum mechanical bounded states. Our result  challenges the practical use of the this type
of method when $\cal H$  fails to be self-adjoint  since, in such cases, we would not have in advance any   reasonable criterion  to
choose the initial eigenstates of $\cal H$ which would correspond to the analytically continued   QNM.
\end{abstract}

\vskip 2cm
%\pacs{ 04.20.-q, 05.20.Dd, 04.40.−b, 47.10.ab}

\date{\today}
\maketitle

\section{Introduction}
\label{Intro}
Usually, the quasinormal  mode (QNM) analysis \cite{Review,Review1} consists in looking  for solutions of the form $\Psi(t,x) = e^{-i\omega t} u(x)$ of an $(1+1)$-dimensional wave equation, with $u(x)$ obeying  a Schroedinger-like equation
\begin{equation}
\label{qnm}
\left(-\epsilon^2\frac{d^2}{dx^2} + V(x)\right) u = \omega^2 u,
\end{equation}
on a certain domain of $\mathbb{R}$, with $\epsilon$ standing for some intrinsic  scale parameter of 
the problem. Typically, as for instance in the problems involving
 asymptotically flat black-hole situations \cite{Review,Review1}, 
the modes $u$ are defined on the entire real line $\mathbb{R}$, which is usually 
assumed to be spanned by a tortoise-coordinate $x$, and $V(x)$ 
is a positive potential barrier, vanishing sufficiently fast when approaching the horizon ($x\to-\infty$)
and the spatial infinity $(x\to\infty)$.  The QNM frequencies are the complex values of
$\omega$ such that the solutions of (\ref{qnm}) behave as outgoing waves at infinity  and ingoing
ones at the horizon, which, according to our definition for $\Psi$, correspond,  respectively, to solutions
such that 
$u \propto e^{ \frac{i\omega x}{\epsilon}}$ for $x\to\infty$ and $u \propto e^{- \frac{i\omega x}{\epsilon} }$ for $x\to-\infty$.
Also according to our definition, the modes will be exponentially suppressed in time, and consequently 
could be interpreted as asymptotically  stable perturbations, if 
$\Im \omega < 0$.

There are several strategies to solve a QNM problem in pratice, see
\cite{Review,Review1} for comprehensive reviews on this vast issue. The   analytical continuation  method, 
introduced decades ago with the pioneering works of Blome, Ferrari, and Mashhoon \cite{ba1,ba2,ba3},  is one of the best options  to have analytical answers and gain some physical insights on the problem. This
method consists basically 
in a formal map between the QNM solutions of (\ref{qnm}) and the bounded states of the quantum mechanical problem
corresponding to  the case of the inverted potential barrier $-V(x)$, which would be governed by the Schroedinger  operator
\begin{equation}
\label{schroed}
{\cal H} =  -\hbar^2\frac{d^2}{dx^2} - V(x) .
\end{equation}
We know that  for $V(x)$  vanishing  sufficiently fast, 
the bounded states of $\cal H$ will decay exponentially, {\em i.e.},
$\psi \propto e^{\mp \sqrt{-E}\frac{x}{\hbar }  }$ for $ x\to\pm \infty $, with
${\cal H}\psi = E\psi$. 
We will adopt  here the analytical continuation  prescription introduced
recently by Hatsuda \cite{borel}, which  is based in the 
  key observation
  that, taking formally $\hbar = i\epsilon$ in (\ref{schroed}), one can map the quantum mechanical bound states of $\cal H$ in the QNM of (\ref{qnm}) with
frequencies formally  given by
\begin{equation}
\label{Hat}
\omega^2 = -E(\hbar = i\epsilon).
\end{equation}
The converse path is also possible, and we will indeed explore it here. 
 Starting with the 
  QNM modes of a potential barrier, one can obtain the associated
  bounded states for the inverted potential Schroedinger operator $\cal H$ by formally setting $\epsilon = -i\hbar $ in (\ref{qnm}) and reversing (\ref{Hat}). 
  Although this construction is not entirely  rigorous and, in particular, we cannot 
   assure that all pertinent
  solutions of the problem may be found by the analytical continuation in the parameters $\epsilon$ and $\hbar$,
  the obtained solutions are certainly valid ones as we  can check    by    direct calculations. 
  
  This kind of analytical continuation approach has proved to be  rather efficient  and, in fact, it has been   used for a vast range
  of QNM analysis recently. In the present paper, we unveil a potential pitfall  when the Schroedinger  operator $\mathcal{H}$ fails to be self-adjoint, which is a rather common situation,
  for instance, when the problem is
  restricted to the half real line
$\mathbb{R}_+$, see \cite{AJP,grif1,GTV,Near} for further references on this kind of
problem. We will exhibit an explicit
example of a problem on $\mathbb{R}_+$ with a well defined and behaved QNM spectrum, but for which the associated Schroedinger
operator $\mathcal{H}$ obtained by the analytical continuation $\epsilon = -i\hbar$ fails to be self-adjoint. Moreover, we show that, although $\mathcal{H}$ admits self-adjoint extensions, the eigenstates corresponding to the analytically continued QNM are not in the domain of 
 any self-adjoint extension of $\mathcal{H}$ and, consequently, they cannot be interpreted as
authentic quantum mechanical bounded states. This means that, if we had chosen  to solve this
problem starting with the quantum mechanical analytically continued $\mathcal{H}$, we would not
have 
any hint or practical prescription
 to choose the initial eigenstates of $\cal H$ which would correspond to the analytically continued QNM. Practically, we would not succeed in applying the analytical continuation method for this kind
of problem.

\section{The QNM  problem}

We will be concerned  here with a specific case where the domain of the QNM is the half real line
$\mathbb{R}_+$. This  
could correspond, for instance, to the case of naked singularities \cite{Naked,Damped}, but certainly there will be other 
 similar cases in distinct situations, see, for instance, \cite{Bin,molina} for   examples
 involving asymptotically AdS spacetimes. We will consider specifically  the infinite potential barrier on  $\mathbb{R}_+$ given by
\begin{equation}
\label{PT}
V(x) = \frac{V_0}{\sinh^2x} + \frac{V_1}{\cosh^2x},
\end{equation}
where $V_0>0$ and $x\in (0,\infty)$,
which is known to be exactly integrable since the seminal works of Poschl and Teller, see for
instance \cite{Bin,molina} and  the references therein. 

We know, on physical grounds, that 
the QNM modes associated the potential (\ref{PT}) should correspond to solutions of (\ref{qnm}) obeying  the boundary conditions
$u(0) = 0$ (infinite impenetrable barrier) and  $u \propto e^{i\omega x}$ for $x\to\infty$ (outgoing
wave in the spatial infinity). 
We will follow
\cite{Bin} closely and introduce  the
variable $z\in (0,1]$ such that $z = 1/\cosh^2x$, 
in terms of which 
 equation (\ref{qnm}) reads
 \begin{equation}
 \label{hyper}
z(1-z) u'' - \left(1 - \frac{3}{2}z \right) u' +\frac{1}{4} \left(
\frac{\omega^2}{z} - \frac{V_0}{1-z} - {V_1}
 \right) u= 0,
 \end{equation}
where the tilde  denotes derivation with respect to $z$. Equation (\ref{hyper}) can be easily
solved in terms of hypergeometric functions and its general solution can be written as combination
of the functions  \cite{Bin}
\begin{equation}
\label{sol}
u = z^\alpha(1-z)^\beta {}_2F_1\left(a,b;c;z\right),
\end{equation}
where
\begin{eqnarray}
a &=& \alpha + \beta + \gamma_+, \\
b &=& \alpha + \beta + \gamma_- , \\
c &=& 2\alpha +1 ,\\
\label{alpha}
\alpha &=& \pm \frac{i\omega }{2}, \\
\label{beta}
\beta &=&    \frac{1}{4}\left(1\pm \sqrt{1+4V_0} \right), \\
\gamma_\pm &=& \frac{1}{4}\left( 1 \pm \sqrt{1-4V_1} \right).
\end{eqnarray}
The general solution  of the  second order linear differential equation (\ref{hyper}) involves two linearly independent functions  
which 
 can be written as a combination of the possible choices
of $\alpha$ and $\beta$ according to (\ref{alpha}) and (\ref{beta}).
However, notice that in 
the spatial infinity limit $x\to\infty$ ($z\to 0^+$), one has $ u = z^\alpha = e^{\mp i\omega x}$   for any choice of $\beta$. Moreover, as one can see from the Wronskian in
the spatial infinity limit, the functions  $u$ corresponding to the two possible choices for
$\alpha$ are linearly independent regardless   the value chosen for $\beta$. Hence, we can
choose without loss of generality the positive sign for $\beta$ for the general solution of (\ref{hyper}). However,  the solution (\ref{sol}) is obtained under the condition of $c \notin \mathbb{Z}$, in fact, the case with $c$ an integer positive or negative will explore later on.

The boundary condition of an outgoing wave at spatial infinity $z\to 0^+$
requires to select only the  $\alpha = -\frac{i\omega }{2}$ mode in the general solution. On the other hand, the  
$x\to 0^+$ limit corresponds to $z\to 1^-$, and thus in order to impose the second QNM boundary condition
$\lim_{x\to 0^+} u =0$, we need to determine the behavior of our solution for $z\to 1^-$. 
By exploring the well known identity 15.3.6 of \cite{Abramowitz},
\begin{widetext}
\begin{eqnarray}\label{expa}
{}_2F_1\left(a,b;c;z\right) &=& \frac{\Gamma(c)\Gamma(c-a-b)}{\Gamma(c-a)\Gamma(c-b)}
{}_2F_1\left(a,b;a+b-c+1;1-z\right) \\ & & + \frac{\Gamma(c)\Gamma(a+b-c)}{\Gamma(a)\Gamma(b)} (1-z)^{c-a-b} 
{}_2F_1\left(c-a,c-b;c-a-b+1;1-z\right), \nonumber 
\end{eqnarray}
\end{widetext}
and that $z=1-x^2 + {\cal O}(x^4)$ for $x\to 0^+$, we have the following behavior for $u(x)$ near $x=0$
\begin{equation}
\label{lim}
u = Ax^{2\beta } \left(1 + {\cal O}(x^2) \right) + Bx^{1-2\beta } \left(1 + {\cal O}(x^2) \right),
\end{equation}
where
\begin{equation}
\label{A}
A = \frac{\Gamma(1-i\omega)\Gamma\left(\frac{1}{2} - 2\beta\right)}{\Gamma\left(
1 - \beta -\gamma_+ - \frac{ i\omega}{2}
 \right) \Gamma\left(
1- \beta  -\gamma_- - \frac{i \omega}{2}
 \right)}
\end{equation}
and
\begin{equation}
\label{B}
B = \frac{\Gamma(1-i\omega)\Gamma\left( 2\beta -\frac{1}{2}\right) }{\Gamma\left(
  \beta + \gamma_+ - \frac{ i\omega}{2}
 \right) \Gamma\left(
 \beta + \gamma_- - \frac{i \omega}{2}
 \right)}.
\end{equation}
Since $\beta > \frac{1}{2}$, the second term in (\ref{lim}) is  always singular for $x\to 0^+$, but it can be indeed
 removed
by exploring the $\Gamma$-function poles in (\ref{B}). Explicitly, we will have $\lim_{x\to 0^+} u =0$
 if the QNM frequencies $\omega$ are such that
\begin{equation}
\label{f3}
\omega_\pm = -i \left( 2n + 1 +  \sqrt{ V_0 + \frac{1}{4}} \pm \sqrt{\frac{1}{4}-V_1}  \right),
\end{equation}
with $n\in \mathbb{Z}_{\ge 0}$ This result coincides with the pertinent limit of the calculations
presented in \cite{Bin}. Notice that, with the choice (\ref{f3}) for the frequencies, the
coefficient $A$ given by (\ref{A}) reads
\begin{equation}
\label{Apm}
A_\pm  = \frac{\left(-1 - \sqrt{ V_0 + \frac{1}{4}}  \right)_n}{\left(-(n+1) - \sqrt{ V_0 + \frac{1}{4}}  \mp
 \sqrt{  \frac{1}{4} - V_1} 
  \right)_n},
\end{equation}
where $( \  )_n$ stands for the usual
falling factorial function
\begin{equation}
(x-1)_n = \frac{\Gamma(x)}{\Gamma(x-n)} .
\end{equation}
Also,  one can see from (\ref{lim}) that 
the QNM are solutions $u(x)$ such that
\begin{equation}
\frac{u(x)}{\sqrt{x}} = Ax^{\delta} \left(1 + {\cal O}(x^2) \right) , 
\end{equation}
and
\begin{equation}
\sqrt{x} u'(x) = \left( \frac{1}{2} + \delta\right) Ax^{\delta} \left(1 + {\cal O}(x^2) \right) ,
\end{equation}
 with the prime now denoting differentiation with respect to $x$,   $A$ given by (\ref{Apm}), and  
 \begin{equation}
 \label{delta}
 \delta = \sqrt{\frac{1}{4}+V_0}.
 \end{equation}
Finally, we have that the QNM are solutions $u(x)$   for which 
\begin{equation}
\label{bc}
\lim_{x\to 0^+} x^{-\delta} \left( \sqrt{x} u'(x) -
\frac{1}{\sqrt x} \left(\frac{1}{2} + \delta \right)   u(x)\right) = 0.
\end{equation}
  This expression for the effective boundary condition obeyed by the QNM 
 at $x=0$ will   be important 
in the self-adjointness analysis of $\mathcal{H}$ in the next section.  

It is worth mentioning
that the simple
exact spectrum (\ref{f3})   illustrates clearly all qualitative behaviors
that QNM can exhibit. 
First, reminding that $V_0>0$, we see that for $V_1 > \frac{1}{4}$, the QNM frequencies will have the
well known form
\begin{equation}
\omega =  \sqrt{  V_1 - \frac{1}{4}} - i\left( 2n + 1 +   \sqrt{ V_0 +\frac{1}{4}} \right),
\end{equation}
denoting that the associated QNM are the usual exponentially suppressed oscillatory modes.
On the other hand, for 
$V_1 \le \frac{1}{4}$, we have the so-called algebraically special QNM, for which the 
associated 
  frequencies are purely imaginary quantities. Notice that if $V_1$ is small enough, one could 
 effectively  attain regimes with   
    $\Im \omega_- > 0$, denouncing the presence of exponentially amplified modes. We will
return to this point in the last section, but we can advance that these unstable modes are 
related to the existence
of a region with $V(x) < 0$   for negative $V_1$.  

If $c=\ell \in \mathbb{Z}_{\geq 1}$ then the second solution of the hypergeometric equation contains a logaritmic branch \cite{Abramowitz}. The latter term is discarded provided it blows up as $z \rightarrow 1^{-}$, that is, one  has to deal with only one solution given by the following expression:
\begin{equation}
\label{sol2}
u_{2} = z^\alpha(1-z)^\beta {}_2F_1\left(a,b,\ell;z\right),
\end{equation}
where this solution ensures the condition of having an out-going wave at infinity as long as $\alpha=\alpha_{-}$. One has to inspect the behavior of (\ref{sol2}) near $x\rightarrow 0$ to see whether or not  $u_{2}$ vanishes. Making use of  (\ref{expa}) and (\ref{lim}) one concludes that the coefficients in (\ref{lim}) are similar to (\ref{A}) and (\ref{B}) but now   the specific condition $c=\ell \in \mathbb{Z}_{\geq 1}$ must be enforced.  The term propotional to $x^{2\beta}$ goes to zero and the Gamma function $\Gamma(1-i\omega)$ is well-behaved provided $\ell$ is not a negative integer. However, the second term in (\ref{lim}) diverges for $2\beta>1$ then it is mandatory that the coefficient $B$ must goes to zero. The latter possiblity is reached when  $2(\beta+\gamma_{\pm})- i\omega=-2n$ with $n \in \mathbb{Z}_{\geq 0}$, which agrees with 
(\ref{f3}) and with the resul reported in \cite{Bin} about the QNMs for pure De Sitter in the massless limit.  Notice also that the spectrum (\ref{f3}) corresponds to the poles of the $\Gamma$-functions 
in the denominator of the coefficient $B$ in (\ref{B}) and, hence, we are
assuming that the respective numerator is regular in these points. Although this is always the
case when  $V_1 > \frac{1}{4}$, we can have situations with \emph{algebraically special modes} for which
the numerators have exactly the same poles of the denominators.  This is the case, for instance,
of $c = 1-i\omega_\pm$ being zero or a negative integer, which implies that the second solution of the hypergeometric equation admits a logaritmic bran   ch and one has to get rid of it \cite{Abramowitz}, yielding to 
\begin{equation}
\label{sol3}
u_{3} = z^{-\alpha}(1-z)^\beta {}_2F_1\left(a-c+1,b-c+1,2-c;z\right),
\end{equation}
where one selects $\alpha=\alpha_{+}$ to reproduce the right behavior at infinity, $u_{3} \propto e^{+i\omega x}$. As  one would expect the condition $c = 1-i\omega_\pm \in \mathbb{Z}_{\leq 0}$ has somekind of impact in the QNMs. In order to see that more clearly, it is convenient to assume that  $V_0 = \mu(\mu+1)$ and $V_1 = \nu(1-\nu )$. Applying the same identity (\ref{expa}) for the hypergeometric function ${}_2F_1\left(a-c+1,b-c+1,2-c;z\right)$ with the new coefficients, one can read off the QNM spectrum  and show that it reduces to
\begin{eqnarray}\label{special1}
\omega_{+} &=& -i(2n + 1  + \mu + \nu),\\  
\omega_{-} &=& -i(2n + 1  + \mu - (\nu-1)),\label{special2}
\end{eqnarray}
with $n\in \mathbb{Z}_{\geq 0}$. Once again, the previous analysis is consistent with the one reported in \cite{Bin}. Eqns. (\ref{special1}-\ref{special2}) tell us that  whenever $\mu \pm \nu$ is a positive integer,
the poles of 
$\Gamma(2-c)=\Gamma(1-i\omega)$  coincide  with those ones in the denominators of $u_3(x)$ when the hypergeometric function ${}_2F_1\left(a-c+1,b-c+1,2-c;z\right)$ is re-written with the help of (\ref{expa}), and consequently we cannot enforce the boundary condition $\lim_{x\to 0^+}u_3 =0$
without having trivially $u_3(x)=0$. There are no
QNMs   in this case, and such a behavior is a remnant of the curious phenomenon of the so-called reflectionless 
eigenstates for the Poschl-Teller potentials in Quantum Mechanics, see \cite{reflectionless}. 

We finish   mentioning that, since $\beta>\frac{1}{2}$,  the QNM limit $u \propto  x^{2\beta}$   
 for $x\to 0^+$   assures that the energy integral
\begin{equation}
{\cal E} = \int \left[ (\partial_t\Psi)^2 +  (\partial_x\Psi)^2 + V(x)\Psi^2\right] dx
\end{equation}
always 
converges for $x\to 0^+$, confirming that the QNM are rather well behaved for the potential barrier (\ref{PT}). 
 
\section{The quantum mechanical problem}

We will address now the following problem: would it be possible  to obtain the QNM of the last section by means of an analytical continuation approach as described in Section \ref{Intro}? As we will see, the answer
is no, at least in the usual sense of associating QNM and bounded states of
the Schroedinger operator (\ref{schroed}) by   an analytical continuation in the parameters
$\epsilon$ and $\hbar$.  
The equivalent quantum mechanical problem can be formulated by means of the formal map
$\epsilon = -i\hbar$ of the first section. Notice that the parameter $\epsilon$ is a ``hidden'' scale
in the QNM problem, we have absorbed it in our calculations so far with the tacit rescalings
$V \leftrightarrow V/\epsilon^2$ and $\omega \leftrightarrow \omega/\epsilon $. Assuming also $\hbar = 1$,
we have that the the analytical continuation will map the QNM $u$ of the potential barrier in eigenstates $\psi$ of
the   potential well corresponding to the inverted potential barrier according to
\begin{equation}
\psi = u \left( V \to - V , \omega \to -i\omega \right).
\end{equation}
The associated eigenvalues (energies) will be given by 
\begin{equation}
E = \omega^2 \left( V \to - V \right),
\end{equation}
and from (\ref{f3})  we have
\begin{equation}
\label{E}
E_\pm = - \left( 2n + 1 +  \sqrt{   \frac{1}{4} - V_0 } \pm \sqrt{\frac{1}{4}+V_1}  \right)^2.
\end{equation}
Such   expression for the energy shows that, in contrast with the QNM analysis of the  potential barrier, now we have 
necessarily to deal with
two qualitatively distinct cases accordingly with the attractiveness of the inverted potential at $x\to 0^+$: $0 < V_0 \le \frac{1}{4}$ and $V_0 > \frac{1}{4}$. This is hardly
a surprise, this class of behavior for an attractive $1/x^2$ potential is well known in the literature, see \cite{grif1,GTV} for instance. 
Moreover, since one can have complex values of $E$ for $V_0>\frac{1}{4}$ or  $V_1< \frac{1}{4}$, the self-adjointness
of   $\cal H$ is clearly  an issue here. 

The self-adjointness analysis of  a Schroedinger operators $\cal H$ always 
starts  \cite{AJP,grif1,GTV,Near}  with the  deficiency indexes $n_+$ and $n_-$,  which stand
here  for the respective   
  dimensions of the so-called deficiency subspaces $N_\pm \subset D({\mathcal{H}^\dagger})$ defined by
\begin{equation}
N_\pm = \left\{ \phi\in  D({\mathcal{H}^\dagger}), \quad {\mathcal{H}}\phi = \pm i \phi\right\},
\end{equation}
where ${\mathcal{H}^\dagger}$ is the adjoint of ${\mathcal{H}}$ and 
$ D({\mathcal{H}^\dagger}) \subset L^2[0,\infty)$ stands for its domain. The deficiency subspaces 
$N_\pm$
 can be constructed  directly form the formulas of the
previous section. For instance, in order to determine $N_+$, one needs to consider the case $\omega^2 = i$, leading 
to two linearly independent solutions which behavior at spatial infinity will be $\phi_\pm \propto e^{\pm i\left(\frac{1+i}{\sqrt 2}\right)x}$. However, only $\phi_+$ is square integrable, and hence we have just
concluded 
that $n_+ \le 1$. Furthermore,  we have also from the results of Section II that, for $x\to 0^+$,
\begin{equation}
\label{psior}
\phi_+ =  \sqrt{x}\left(  A_+ x^{ \delta} + B_+ x^{- \delta}  \right),
\end{equation}
with $\delta $  given by the substitution $V_0 \to -V_0$ in (\ref{delta})  and  $A_+$ and $B_+$ 
obtained by setting $\omega = \frac{1+i}{\sqrt 2}$
in the expressions (\ref{A}) and (\ref{B}), respectively. Notice that (\ref{psior}) is square integrable for  $x\to 0^+$ and no other restriction is necessary on $\phi_+$. We have just  established that $n_+ = 1$. Since similar results hold for
$N_-$,   we have that ${\mathcal{H}}$ is not self-adjoint, but the 
von Neumann formulas \cite{GTV} assure  that ${\mathcal{H}}$ admits a one-parameter family of self-adjoint extensions.  In order to construct such extensions, we need the the function
$\phi_-$, {\em i.e.}, the square integrable solution for $ {\mathcal{H}}\phi = -i \phi$. Its behavior
for $x\to 0^+$ is
\begin{equation}
\label{psiorm}
\phi_- = \sqrt{x}\left( A_-  x^{ \delta} + B_- x^{- \delta}  \right),
\end{equation}
with the constants $A_-$ and $B_-$ defined as in (\ref{psior}), but now for $\omega = \frac{i-1}{\sqrt 2}$.
The self-adjoint extensions of $\mathcal H$ are determined by the boundary conditions on 
$\psi\in  D({\mathcal{H}^\dagger}) $ such
that
\begin{equation}
\label{ext}
\left\langle \phi  , \mathcal {H} \psi \right\rangle  - 
\left\langle  \mathcal {H} \phi  ,  \psi \right\rangle = \lim_{x\to 0^+}[ 
\bar\phi (x) \psi'(x) -{\bar\phi  }'(x)  \psi(x)]  = 0,
\end{equation}
where the inner product is the  usual $L ^2[0,\infty)$ product and $\phi$ is a linear combination of $\phi_+$ and $\phi_-$, which we write without loss of generality as
\begin{equation}
\phi = \phi_+ + \bar\lambda \phi_-,
\end{equation}
where $\lambda$ is a free complex parameter. From (\ref{ext}), we have that 
the self-adjoint extensions of $\mathcal{H}$ will be given by the following boundary conditions on
the eingenstates $\psi$
\begin{widetext}
\begin{equation}
\label{boundary}
\lim_{x\to 0^+}  \left( 
\sqrt{x} \left[\mathcal{A}_\lambda x^{\bar \delta }  + \mathcal{B}_\lambda x^{-\bar \delta }\right] 
\psi'(x)  -
\frac{1}{\sqrt{x}} \left[ \left(\frac{1}{2}+\bar\delta\right)  \mathcal{A}_\lambda x^{\bar \delta }  +  \left(\frac{1}{2}-\bar\delta\right)\mathcal{B}_\lambda x^{-\bar \delta } \right]\psi(x) \right) = 0,
\end{equation}
\end{widetext}
where
\begin{equation}
\mathcal{A}_\lambda = \bar A_+ + \lambda \bar A_-
\end{equation}
and 
\begin{equation}
\mathcal{B}_\lambda = \bar B_+ + \lambda \bar B_-.
\end{equation}
For each value of the free parameter $\lambda$, we have a distinct self-adjoint extension of the
Schroedinger operator  $\mathcal{H}$ characterized by the boundary condition (\ref{boundary})
at $x=0$.  Notice that (\ref{boundary}) coincides wit the usual
self-adjoint boundary condition of $1/x^2$ attractive potentials, the so-called Calogero problem, see \cite{grif1,GTV}.

Now, let us consider the QNM boundary condition (\ref{bc}),
which the analytically continued eigenstates would also be expected to obey in the quantum mechanical problem.
   We have two qualitatively distinct cases. 
 Let us consider first the case of real nonzero 
$\delta$, {\em i.e.}, $ 0 < V_0 < \frac{1}{4}$. The QNM boundary condition requires  a value for $\lambda$ 
such that $\mathcal{A}_\lambda = 0$
in order to select the $x^{-\delta}$ terms in (\ref{boundary}). However, in this case one has
\begin{equation}
\label{qbc}
\lim_{x\to 0^+} x^{-\delta} \left( \sqrt{x} u'(x) -
\frac{1}{\sqrt x} \left(\frac{1}{2} - \delta \right)  u(x)\right) = 0,
\end{equation}
which clearly does not coincide with (\ref{bc}), meaning that no self-adjoint extension of $\mathcal H$ 
can accommodate the QNM boundary condition for the case of real nonzero $\delta$. 
The so-called critical case $\delta=0$ could be 
analyzed in a similar way, but in this case the deficiency subspaces $N_\pm$ are generated by
the solutions
\begin{equation}
\phi_\pm = \sqrt{x}\left( A_\pm    + B_\pm \ln  x   \right),
\end{equation}
 see   \cite{grif1,GTV} for further details on this particular case of the
 $1/x^2$ potential.  
The situation for
imaginary $\delta$ is analogous. We need to have $\mathcal{B}_\lambda =0$, 
to select the $x^{\bar\delta}$ terms, but we will end up with  
  (\ref{qbc}) once more, implying again that no self-adjoint extension of $\mathcal H$ 
will be able to accommodate the QNM boundary condition for the case of imaginary $\delta$ as well. 

In summary, although $\mathcal{H}$ fails to be self-adjoint
in the quantum mechanical problem corresponding to the inverted potential
(\ref{PT}), it admits self-adjoint extensions. However,
 the eigenstates corresponding to the analytically continued QNM are not in the domain of 
 any self-adjoint extension of $\mathcal{H}$ and, consequently, they cannot be interpreted as
authentic quantum mechanical bounded states. In particular, they can fail to have
real eigenvalues, see (\ref{E}),  and to be 
square-integrable.

\section{Final remarks} 

\begin{figure}[h]
\includegraphics[scale=0.5]{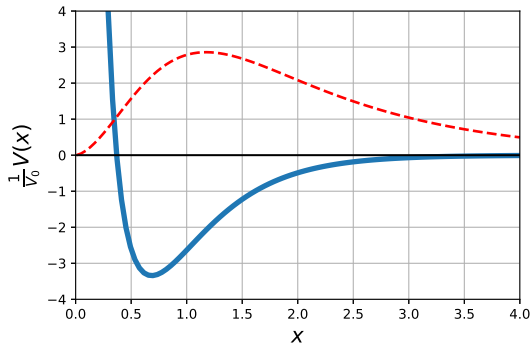} 
\caption{\label{fig1}\footnotesize
Illustration of an unstable algebraically special QNM. 
 Solid blue line: Potential barrier (\ref{PT}) for $V_0/V_1 = -\frac{1}{8}$.
Red traced line: the unstable mode (no scale) corresponding to the $n=0$ case in the   frequency  $\omega_-$ in (\ref{f3}). 
}
\end{figure}

Let us return to the unstable algebraically special modes presented in (\ref{f3}) for negative $V_1$. Although the 
behavior of the potential barrier (\ref{PT}) depends only on $V_0$ for $x\to 0^+$, the situation 
for $x\to\infty$ is clear different, since we have
\begin{equation}
V(x) = \frac{4(V_0 + V_1)}{e^{2x}}\left( 1+ \mathcal{O}(e^{-2x})\right)
\end{equation}
in this limit. As one can see, for $V_0+V_1 < 0$,   the potential $V$ approaches zero for $x\to\infty$ from below and,
hence, it must have a minimum somewhere in $\mathbb{R}^+$. It is indeed easy to locate this minimum, it corresponds
to the point $x=x^*$ such that
\begin{equation}
\tanh^4x^* = - \frac{V_0}{V_1},
\end{equation}
for which we have simply
\begin{equation}
V(x^*) = -\left( \sqrt{V_0} - \sqrt{-V_1} \right)^2,
\end{equation}
remembering that $V_0>0$ and $V_1 < - V_0$. 
Fig. \ref{fig1} illustrates  a typical 
 unstable algebraically special QNM. The curious observation is that 
 the unstable QNM with a purely imaginary QNM frequency  (\ref{f3}) corresponds, in fact, to a bounded state
 of the   potential barrier! The deficiency indexes for  the Schroedinger operator associated with
 the original potential (\ref{PT}) are $n_+ = n_- = 0$ since no solution of 
 $\mathcal{H}u = \pm i u$ is square integrable, implying that $\mathcal{H}$ for the case of a repulsive
 potential is self-adjoint and, in particular, it may admit bounded states as that one 
 depicted in Fig. \ref{fig1}.
 
\section{summary} 

We have shown that the QNM solutions $u$ of the potential barrier $V$ given by (\ref{PT}) can be analytically 
continued in eigenstates $\psi$ of the Schroedinger operator $\mathcal{H}$ associated with the inverted
potential $-V$. However, the eigenstates $\psi$ are not in the domain of any
self-adjoint extension of $\mathcal{H}$, challenging the practical use of the analytical continuation
method for this kind of potential barrier. If we had tried to solve this problem starting
with the states of $\mathcal{H}$, 
we would not have any physical or mathematical criterion to select the states $\psi$ which would
given origin to the QNM $u$ via  the analytical continuation. Furthermore, if we have tried starting with a genuine quantum state of $\mathcal{H}$, after the analytical continuation we would get a solution of the QNM problem which does not obey the correct boundary conditions. In other words, this result poses a serious challenge on how to extend the analytic continuation method in several spacetimes (naked singularity, pure De sitter, etc) to obtain the proper QNMs. Whether the analytic continuation approach can be applied to  other more complex scenarios or not is something that deserves to be explored in the future. For instance, one may wonder about its  applicability  in the case of a Dirac field propagating on those backgrounds \cite{Bin},\cite{kermions}. The  so-called PT-symmetric extensions of the Schroedinger operator (see, for instance, \cite{Bender} and references therein) and the pseudospectrum
 of  non-selfadjoint operators \cite{pseudo}  are also interesting points to be explored 
in the context of our QNM analysis.
 
\hspace{2cm} 

%%%%%%%%%%%%%%%%%%%%%%%%%%%%%%%%%%%%%%%%%%%%%%%%%%
\acknowledgments
%%%%%%%%%%%%%%%%%%%%%%%%%%%%%%%%%%%%%%%%%%%%%%%%%

We thank Jorge Zanelli for several  discussions.
J.C.F. is supported by Conselho Nacional de Desenvolvimento Cient\'ifico e Tecnol\'ogico (CNPq, Brazil) and  Funda\c{c}\~ao de Amparo \'a Pesquisa e Inova\c{c}\~ao Esp\'irito Santo (FAPES, Brazil). M.G.R is supported by FAPES/CAPES grant under the PPGCosmo Fellowship Program. A.S. is partially supported by Conselho Nacional de Desenvolvimento Cient\'ifico e Tecnol\'ogico (CNPq, Brazil).
\vspace{0.6cm}
%%%%%%%%%%%%%%%%%%%%%%%%%%%%%%%%%%%%%%%%%%%%%%%%%%%%%%%%%%%%%%%%%%%%%%%%%%%%%%%

%%%%%%%%%%%%%%%%%%%%%%%%%%%%%%%%%%%%%%%%%%%%%%%%%%%%%%%%%%%%%%%%%%%%%%%%%%%%%%%%%%%%%%%%%%%%%%%%%%%%%%%%%%%%%%%%%%%%%%%%%%%%
\end{document}